\documentclass[12pt,3p,,preprint]{elsarticle}
\usepackage[LGR,T1]{fontenc}
\usepackage[utf8]{inputenc}
\usepackage{xcolor}
\usepackage{textcomp}
\usepackage{amsmath}
\usepackage{amssymb}
\usepackage{cancel}
\usepackage{stmaryrd}
\usepackage{graphicx}
\usepackage{subscript}

\makeatletter

\ProvideTextCommand{\~}{LGR}[1]{\char126#1}

\if@twocolumn\PassOptionsToPackage{switch}{lineno}\else\fi

\usepackage{tabulary}
\usepackage{xcolor}\usepackage{amsfonts}

\let\save@ps@pprintTitle\ps@pprintTitle
\def\ps@pprintTitle{\save@ps@pprintTitle\gdef\@oddfoot{\footnotesize\itshape \null\hfill\today}}
\def\hlinewd#1{%
  \noalign{\ifnum0=`}\fi\hrule \@height #1%
  \futurelet\reserved@a\@xhline}

\AtBeginDocument{\ifNAT@numbers \biboptions{sort&compress}\fi}

\usepackage{ifluatex}
\ifluatex
\usepackage{fontspec}
\defaultfontfeatures{Ligatures=TeX}
\usepackage{unicode-math}
\unimathsetup{math-style=TeX}
\else 
\fi 
\ifluatex\else\fi

\usepackage{url}
\usepackage{multirow}\usepackage{morefloats}\usepackage{floatflt}\usepackage{tfrupee}

\AtBeginDocument{\@ifpackageloaded{textcomp}{}{\usepackage{textcomp}}}

\usepackage{colortbl}
\usepackage{xcolor}
\usepackage{pifont}
\usepackage[nointegrals]{wasysym}
\def\mcWidth#1{\csname TY@F#1\endcsname+\tabcolsep}

\def\cAlignHack{\rightskip\@flushglue\leftskip\@flushglue\parindent\z@\parfillskip\z@skip}
\def\rAlignHack{\rightskip\z@skip\leftskip\@flushglue \parindent\z@\parfillskip\z@skip}

\usepackage{ifxetex}
\ifxetex\else\if@twocolumn\usepackage{dblfloatfix}\fi\fi

\AtBeginDocument{
\expandafter\ifx\csname eqalign\endcsname\relax
\def\eqalign#1{\null\vcenter{\def\\{\cr}\openup\jot\m@th
  \ialign{\strut$\displaystyle{##}$\hfil&$\displaystyle{{}##}$\hfil
      \crcr#1\crcr}}\,}
\fi
}

\AtBeginDocument{%
  \@ifpackageloaded{endfloat}%
   {\renewcommand\efloat@iwrite[1]{\immediate\expandafter\protected@write\csname efloat@post#1\endcsname{}}}{\newif\ifefloat@tables}%
}%

\def\BreakURLText#1{\@tfor\brk@tempa:=#1\do{\brk@tempa\hskip0pt}}
\let\lt=<
\let\gt=>
\def\processVert{\ifmmode|\else\textbar\fi}

\@ifundefined{subparagraph}{
\def\subparagraph{\@startsection{paragraph}{5}{2\parindent}{0ex plus 0.1ex minus 0.1ex}%
{0ex}{\normalfont\small\itshape}}%
}{}

\newcommand{\role}[1]{\unskip}
\newcommand{\aucollab}[1]{\unskip}
  
\@ifundefined{tsGraphicsScaleX}{\gdef\tsGraphicsScaleX{1}}{}
\@ifundefined{tsGraphicsScaleY}{\gdef\tsGraphicsScaleY{.9}}{}
\def\checkGraphicsWidth{\ifdim\Gin@nat@width>\linewidth
	\tsGraphicsScaleX\linewidth\else\Gin@nat@width\fi}

\def\checkGraphicsHeight{\ifdim\Gin@nat@height>.9\textheight
	\tsGraphicsScaleY\textheight\else\Gin@nat@height\fi}

\def\fixFloatSize#1{}
\let\ts@includegraphics\includegraphics

\def\inlinegraphic[#1]#2{{\edef\@tempa{#1}\edef\baseline@shift{\ifx\@tempa\@empty0\else#1\fi}\edef\tempZ{\the\numexpr(\numexpr(\baseline@shift*\f@size/100))}\protect\raisebox{\tempZ pt}{\ts@includegraphics{#2}}}}

\AtBeginDocument{\def\includegraphics{\@ifnextchar[{\ts@includegraphics}{\ts@includegraphics[width=\checkGraphicsWidth,height=\checkGraphicsHeight,keepaspectratio]}}}

\DeclareMathAlphabet{\mathpzc}{OT1}{pzc}{m}{it}

\def\URL#1#2{\@ifundefined{href}{#2}{\href{#1}{#2}}}

\def\UrlOrds{\do\*\do\-\do\~\do\'\do\"\do\-}%
\g@addto@macro{\UrlBreaks}{\UrlOrds}

\@ifundefined{quoteAttrib}{}{}

\@ifundefined{titlequoteAttrib}{}{}

\newenvironment{title-quote}{\list{}{\fontsize{10pt}{12pt}\selectfont\leftmargin.5in\itshape\rightmargin\leftmargin}%
  \item\relax}{\endlist}

\usepackage{float}

\makeatother

\begin{document}
\begin{frontmatter}

\title{Faceting diagram for Ag segregation induced nanofaceting at an asymmetric
Cu tilt grain boundary}

\author[affd29b90a2311c44f23c88541eb15cf46b]{Nicolas J. Peter\corref{contrib-246c5b1a11899cc6f22c947ba0da0831}}

\ead{peter@mpie.de}

\author[affd29b90a2311c44f23c88541eb15cf46b]{Maria J. Duarte}

\ead{duarte@mpie.de}

\author[affd29b90a2311c44f23c88541eb15cf46b]{Christoph Kirchlechner}

\ead{kirchlechner@mpie.de}

\author[affd29b90a2311c44f23c88541eb15cf46b]{Christian H. Liebscher\corref{contrib-246c5b1a11899cc6f22c947ba0da0831}}

\ead{liebscher@mpie.de}

\author[affd29b90a2311c44f23c88541eb15cf46b]{Gerhard Dehm}

\ead{dehm@mpie.de}

\cortext[contrib-246c5b1a11899cc6f22c947ba0da0831]{Corresponding author.}

\address[affd29b90a2311c44f23c88541eb15cf46b]{ Max-Planck-Institut f{ü}r Eisenforschung GmbH\unskip, Max-Planck-Strasse
1\unskip, 40237 D{ü}sseldorf\unskip, Germany}
\begin{abstract}
In this work, we experimentally establish
the isothermal nanofacet evolution at an asymmetric tilt grain boundary
in the Cu-Ag system using a diffusion couple approach. We investigate
the nanofacet formation along the grain boundary in dependence of
the Ag solute excess concentration. The initial grain boundary dissociates
into asymmetric Ag-lean segments and Ag-rich symmetric (210) segments.
Increasing Ag excess leads to an increase in Ag-rich facet segment
length, while the length of the asymmetric facets remains constant.
From this, we construct a grain boundary nanofaceting diagram deduced
from our experiments relating local atomic structure, overall inclination
and Ag solute excess.
\end{abstract}
\end{frontmatter}

Grain boundaries (GBs) are interfaces
separating adjoining crystals with different misorientation. They
are amongst the most important material defects and their associated
structure and properties determine their influence on the behavior
of polycrystalline materials. In early work, Gibbs established that
interfacial transitions can be described by equilibrium thermodynamics
\cite{gibbs1945collected} and Hart later proposed that transitions
at GBs are possible and referred to them as ``two-dimensional phase
transitions'' \cite{hart1968two}. The thermodynamic description
of these transitions was extended by Cahn and Rottman \cite{cahn1982transitions,rottman1988theory}
who categorized them into faceting and congruent (all 5 degrees of
freedom remain constant) transition\textcolor{black}{s. Both types,
non-congruent (faceting) and congruent transitions ar}e referred to
as GB phase transformations \cite{o2018grain} or synonymously GB
complexion transitions \cite{dillon2007complexion,kaplan2013review}.
However, these transitions are typically studied in perfectly symmetric
and planar GBs \cite{duscher2004bismuth,meiners2020observations,parajuli2019misorientation}
and mostly derived from indirect experimental measurements \cite{divinski2012diffusion,prokoshkina2017experimental}
or atomistic simulations \cite{frolov2013effect}. Instances of how
complex such GB transitions can be and how factors like temperature
and chemistry can influence a GB's structure were recently demonstrated
by scanning transmission electron microscopy (STEM). Hu et al. found
a temperature dependent segregation behavior including a first order
congruent transition at low temperatures in a planar Au-doped Si GB
using atomistic calculations \cite{hu2019first}. Very recently, a
complex co-segregation pattern including segregation induced symmetry
changes and order losses at a planar WC grain boundary was discovered
\cite{luo2019highly}. A comprehensive review on evidence of grain
boundary transitions can be found in a recent review article \cite{cantwell2020grain}.

Already Cahn discussed that most GB transitions involve faceting \cite{cahn1982transitions}
and such transitions have been observed experimentally in many materials
\cite{donald1979grain,choi2001temperature,koo2001the}. Faceting-roughening
phenomena of GBs were recently summarized by Straumal et al. for a
great amount of different grain boundary types and in different material
systems \cite{straumal2016grain}. The majority of experimental and
simulation work established a clear dependence of faceting-defaceting
transitions on temperature in pure metal GBs \cite{barg1995faceting,bowers2016step,brown2007dissociation,choi2001temperature}.
Some work also focused on how structural defects of a grain boundary
(GB dislocations) may influence faceting by pinning or dragging facets
during defect motion \cite{shibata2003grain}. \textcolor{black}{Transitions
influenced by chemical stimulus, for instance by segregation, are
much less understood} in the case of GB faceting as compared to the
segregation behavior of planar boundaries. Changing the chemistry
at GBs by adsorption of solute excess significantly changes the energy
of such boundaries and consequently structural transitions are likely
to occur. Early work showed structural transitions at a low angle
grain boundary in the Fe-Au system \cite{sickafus1987grain}. Faceting
transitions were frequently observed in the last decades in the Cu-Bi
binary system since Bi promotes GB embrittlement \cite{sigle2002correlation}.
Experimentally, only one carefully performed \textit{in situ} experiment
conducted by Ference and Baluffi unambiguously demonstrated that reversible
faceting-defaceting can be induced by solute adsorption \cite{ference1988observation}.
However, faceting occurred on a scale of several tens to hundreds
of nanometers and the characterization of the atomic GB structure
and chemistry was not possible at that time. Four decades ago Loier
and Boos demonstrated already how small additions of Bi induce striations
and faceting of Ni grain boundaries \cite{loier1981striation}. Yu
et al. recently found distinct segregation patterns of Bi solutes
at a microfaceted general GB in Ni \cite{yu2017segregation} after
having shown that such patterns may be the underlying reason for detrimental
effects of Bi segregation on mechanical properties in Ni \cite{luo2011role}.
Our recent work combining atomic resolution imaging and atomistic
modelling demonstrated that Ag segregation can act as chemical trigger
promoting a nanofaceting transition of an asymmetric tilt grain boundary
in Cu \cite{peter2018segregation}. However, temperature \cite{frolov2013structural},
pressure \cite{harris1999computer} or solute excess concentration
\cite{frolov2015segregation,ference1988observation} strongly influence
GB transitions \cite{frolov2012thermodynamics} and their concentration
dependence is usually established by atomistic modeling. 

Here, we report the experimental assessment of GB plane and faceting
evolution in dependence on the Ag solute excess concentration at an
initially asymmetric tilt grain boundary in Cu. The concentration
dependence was observed using atomic resolution STEM in combination
with energy dispersive X-ray spectroscopy (EDS) and atom probe tomography
(APT) in close vicinity to the interface of a Cu-Ag diffusion couple.
A Ag reservoir was sputter deposited on top of a bicrystal containing
the GB to create the diffusion couple. For consistency, we refer to
the surface of the bicrystal (the interface to the Ag reservoir) as
surface\textcolor{black}{\emph{.}}\textcolor{black}{{} }

\textit{\textcolor{black}{\emph{A pure Cu reference boundary of the
bicrystal's top surface before Ag deposition was observed at three
distinct magnifications (Fig. 1b).}}}\textit{\textcolor{black}{{} }}\textit{\textcolor{black}{\emph{Temperature
induced faceting of that boundary was already excluded in our previous
work \cite{peter2018segregation}.}}}\textcolor{black}{{} The boundary
appears to be entirely flat for at least 13 \textmu m from the}\textit{\textcolor{black}{{}
}}\textit{\textcolor{black}{\emph{top}}}\textcolor{black}{\emph{ }}\textcolor{black}{surface
into the bulk as determined by STEM, even at higher magnification.
Atomic resolution imaging reveals the local atomic arrangement and
although some sub-nanometer roughness can be associated to the asymmetric
GB}\textcolor{black}{\emph{;}}\textcolor{black}{{} we refer to this
boundary as flat reference boundary. The reference GB was determined
to be close to an asymmetric 34°{[}001{]} (920)/(110) boundary}\textcolor{black}{\emph{,
}}\textit{\textcolor{black}{\emph{which is in close agreement to our
previously investigated GB showing the Ag induced nanofaceting \cite{peter2018segregation}}}}\textcolor{black}{.
After Ag diffusion, the overall macroscopic grain boundary (observed
at low magnification) is no longer straight and a seemingly continuous
curvature from the bulk to the surface was induced. A schematic representation
of the grain boundary after segregation is given in Fig. 1a) and an
}\textit{\textcolor{black}{\emph{experimental HAADF STEM}}}\textcolor{black}{{}
micrograph of a part of the grain boundary }\textit{\textcolor{black}{\emph{close
to the surface}}}\textcolor{black}{\emph{ }}\textcolor{black}{at low
magnification in Fig. 2. Micrographs were then acquired at certain
locations and at high magnifications along the segregated GB to study
the overall plane evolution connected with the nanofacet formation
and the associated Ag excess. Representative locations (}\textit{\textcolor{black}{\emph{see
Fig. 2 positions}}}\textit{\textcolor{black}{{} }}\textcolor{black}{4
to 1}\textit{\textcolor{black}{, }}\textit{\textcolor{black}{\emph{from
Cu(Ag) GB close to the}}}\textit{\textcolor{black}{{} }}\textit{\textcolor{black}{\emph{surface}}}\textcolor{black}{)
are marked in the overview image and are presented in more detail
below. We observe a gradual transition from a flat segregated boundary
(1 in Fig. 2, compare }\textit{\textcolor{black}{\emph{STEM-EDS results}}}\textcolor{black}{),
in close resemblance to the Ag-free reference, to a symmetric \ensuremath{\sum}5
(210) GB (4 in Fig. 2) as further indicated by the adoption of the
kite motif \cite{peter2018segregation,frolov2013effect} in the enlarged
area next to the micrograph of position 4.} \textcolor{black}{The
intermediate GB adopts a nanofaceted state containing Ag-rich, symmetric
\ensuremath{\sum}5 (210) segments and newly formed, asymmetric and
Ag-lean segments with a GB plane close to (320)/(100).} Almost all
kite tip positions are occupied by Ag (bright columns in HAADF STEM).
\textcolor{black}{In addition, we occasionally observe image contrast
indicating Ag occupation of non-kite tip positions at the GB, as well.
Therefore it is reasonable to believe that the symmetric GB close
to the surface is covered close to a monolayer of Ag. }The symmetric
\ensuremath{\sum}5 (210) GB plane was measured to be $\sim$ 20º inclined
with respect to the non-faceted, asymmetric (920)/(110) reference
boundary plane. 

The precise solute excess concentration and localized elemental distribution
were determined along the boundary by APT and STEM-EDS, respectively.
An APT specimen extracted close to the sample surface (Tip 1, \textit{\textcolor{black}{\emph{position
4 in Fig. 2}}}) was found to have a peak concentration of $\sim$3.6
at. \% Ag (black line) as determined from the integrated line profiles
across the GB (Fig. 3a, upper graph). A specimen extracted from below
the sample surface (Tip 2, \textit{\textcolor{black}{\emph{at position
3 in Fig. 2}}}) only contained a peak solute excess of $\sim$2.4
at. \% Ag (Fig. 3a, lower graph). \textit{\textcolor{black}{\emph{A
depletion zone is visible for the Ag concentration next to the enriched
GB in both specimen. In the case of the less segregated GB (Tip 2),
the Ag concentration recovers towards the outer APT needle's surface
and reaches saturation at a level of about 1.3 at. \% at about 25
nm from the GB. For Tip 1, the depletion extends much further and
saturation is not reached inside the specimen volume. From the course
of the concentration profile we expect a saturation level of around
1.7 at. \% at about 45 nm from the GB deduced from a second order
polynomial fitting. To determine the respective bulk Ag concentration
more reliable, we created cylinders of the same dimensions (25 nm
x 10 nm x 40 nm) inside the two volumes and placed them parallel to
the GB at a fixed distance of 20 nm (in saturation level of Tip 2).
Bulk concentrations of 1.43}}}\textcolor{black}{\emph{ }}\textcolor{black}{\textpm{}
0.015}\textcolor{black}{\emph{ }}\textit{\textcolor{black}{\emph{at.
\% and 1.23}}}\textcolor{black}{\emph{ }}\textcolor{black}{\textpm{}
0.011 a}\textit{\textcolor{black}{\emph{t. \% were extracted for Tip
1 and Tip 2, respectively.}}}\textit{\textcolor{black}{{} }}\textit{\textcolor{black}{\emph{The
precise location of the extracted tips within a few \textmu m is hardly
possible to be determined after FIB milling preparation. The bulk
Ag concentrations appear reasonable considering the low Ag solubility
according to the equilibrium phase diagram \cite{subramanian1993ag}}}}.
In addition, Ag density maps of thin slices \textcolor{black}{(1 nm
thickness)} through the reconstructed volumes of these two APT tips
- one from a surface-near region (Tip 1) and one extracted from\textcolor{black}{{}
an approximate location 3 in Fig. 2 }(Tip 2) - revealed a non-faceted,
Ag-rich boundary close to the surface and further towards the bulk
a faceted, Ag-rich boundary. This transition agrees well with our
experimental STEM micrographs. Across these two boundaries we calculated
the solute excess concentration following our previously applied approach
\textcolor{black}{\cite{peter2018segregation,felfer2012quantitative}}
and revealed \textit{\textcolor{black}{\emph{for Tip 1}}} a Ag excess
concentration of 4.96 at/nm\textsuperscript{2}, while the faceted
GB\textcolor{red}{\emph{ }}\textit{\textcolor{black}{\emph{(Tip 2)}}}\textcolor{red}{\emph{
}}showed an average excess concentration of 3.39 at/nm\textsuperscript{2}.
\textit{\textcolor{black}{\emph{Based on the latter, we chose a maximum
value for the color bar to qualitatively best show the faceted GB
state.}}} These results appear reasonable, as STEM results revealed
preferential segregation to symmetric \textcolor{black}{\ensuremath{\sum}5}
(210) GB portions, which dominate at the surface, but contribute less
for deeper measurement locations\textit{\textcolor{violet}{{} }}\textcolor{black}{into
the bulk due to the increasing incorporation of asymmetric Ag-lean
segments}. In addition, the diffusion gradient from the surface into
the bulk reduces the Ag excess concentration at the GB, which therefore
appears to be coupled with the facet lengths. The excess of the non-faceted,
\textit{\textcolor{black}{\emph{symmetric}}} \textcolor{black}{\ensuremath{\sum}5}
(210) boundary close to the surface is just slightly below the monolayer
solute excess of roughly 6 at/nm\textsuperscript{2}, \textit{\textcolor{black}{\emph{which
is calculated by the number of atoms on a (210) plane unit area}}}.
However, STEM micrographs at the very surface revealed some areas
in which two bright atomic columns appear next to each other, which
is why we assume to have at least monolayer coverage at the very surface
of the diffusion couple, i.e. close to the Ag reservoir. The drawback
of such compositionally sensitive APT measurements is the lack of
atomically resolved structural features in the analyzed volume. Consequently,
STEM-EDS was used in order to correlate the occurrence of the three
structural states observed with their solute excess\textcolor{black}{{}
}at different locations along the segregated boundary\textcolor{black}{.
Integrated intensity line profiles for the Ag-L}\textsubscript{$\alpha$}\textcolor{black}{{}
peak were extracted from EDS maps at the investigated locations and
three representative profiles (close to positions 1, 2 and 4) for
the three observed GB states are plotted after applying of a moving
average algorithm for visibility in Fig. 3b) normalized to the background
intensity. The inset shows EDS spectra integrated in a region of about
10 x 15 nm}\textsuperscript{\textcolor{black}{2}}\textcolor{black}{{}
of the expected Ag-L}\textsubscript{$\alpha$}\textcolor{black}{{}
peak for a GB containing area and an area inside one grain for comparison.
We would like to point out that EDS maps were acquired in zone axis
orientation immediately after image acquisition to avoid tilting and
loss of spatial correlation. In this case, quantification of the EDS
data is no longer applicable through the Cliff-Lorimer method and
intensive simulation work would have to be performed to account for
effects like beam spreading and electron channelling \cite{kothleitner2014quantitative}.
Therefore, we do not give absolute concentration values from EDS but
normalize our results to the concentration expected for monolayer
coverage by integrating the peak intensity and collapsing it to a
GB width of 0.12 nm, as this width provides a Ag reference concentration
of 100 \% for the data set closest to the surface. This assumption
is based on our experimental }\textit{\textcolor{black}{\emph{HAADF
STEM results. STEM-EDS analyses}}}\textcolor{black}{{} show a clear
increase of Ag from the bulk towards the surface. }

\textcolor{black}{The combination of structural and compositional
characterization of the observed faceting evolution enables a comprehensive
view on the evolution of individual facet segments. In fact, we composed
a segregation-induced GB faceting diagram for the investigated GB
under the chosen diffusion conditions (Fig. 4). Plotting the facet
segment length against the relative Ag monolayer coverage (Fig. 4a)
as determined by EDS along the GB (Fig. 3b) reveals a concentration
window for faceting to occur. While the asymmetric facet segment length
was found to be independent of the Ag excess and remained at a length
$l$ of about 1 nm, the symmetric facet segments almost linearly grew
with increasing excess concentration starting from 1 nm (R-square
of 0.96). Therefore, the reason for an increasing global GB inclination
lies in growing symmetric facet segments. Thus there are three clearly
distinguishable GB states as a function of overall Ag excess solute:
(i) non-faceted, asymmetric and Ag-lean (Figure 2-1), (ii) preferentially
Ag segregated and nanofaceted (Figure 2-2 and 2-3) as well as (iii)
non-faceted, symmetric and Ag-rich (Figure 2-4). }

\textcolor{black}{The present results confirm our previous findings
\cite{peter2018segregation} that after Ag segregation induced facet
formation there is preferential segregation occurring to only the
symmetric facet segments, while the asymmetric segments remain Ag-lean.
To make a better connection between facet segment length, Ag segregation
and especially overall GB inclination, we plotted the relative Ag
excess concentration as obtained by EDS under the assumption of full
monolayer coverage for the symmetric \ensuremath{\sum}5 (210) GB close
to the surface (Fig. 4b). At this point, the boundary was about 20°
inclined from the asymmetric reference GB. }\textit{\textcolor{black}{\emph{To
describe the relation between Ag concentration and facet length, we
normalize the ratio of the asymmetric segment length (ASL) and the
symmetric facet segment length (SSL) using the ``facet segment ratio''
parameter FSR. FSR is calculated as $FSR=SSL/(ASL+SSL)$.}}}\textcolor{black}{{}
Consequently, the}\textcolor{red}{\emph{ }}\textit{\textcolor{black}{\emph{Ag-rich,
symmetric}}}\textcolor{black}{{} \ensuremath{\sum}5 (210) GB close to
the surface does not show any faceting and FSR = 1 (ASL = 0). The
purely asymmetric reference GB at inclination 0° has consequently
no symmetric segments and thus FSR = 0. The ASL was determined above
to be }\textit{\textcolor{black}{\emph{approximately}}}\textcolor{red}{\emph{
}}\textcolor{black}{constant with a length of }$\sim$\textcolor{black}{1
nm }\textit{\textcolor{black}{\emph{for the faceted GB}}}\textcolor{black}{.
Under the assumption of full monolayer coverage for FSR = 1 supported
by STEM observations, we drew a dashed line (red) to indicate an anticipated
linear decrease of Ag excess with a linearly decreasing amount of
the symmetric segment fraction. Our experimental results are given
as individual data points and are linearly fitted (white) with a R-square
value of 0.93. Indeed, our results follow the anticipated, ideal line
in reasonable agreement (compare red and white lines), which leads
us to the conclusion that for the nanofaceted boundary portion symmetric
facet segments always are at least close to monolayer coverage, while
it indicates that asymmetric segments of the nanofaceted grain boundary
section are close to being Ag-free. This extends our previous findings
deduced from a GB faceted at a single Ag concentration to the entire
faceting range. No faceting was observed below FSR = 0.5, which means
below these equisized nanofacets, faceting disappears and a purely
asymmetric GB takes up the Ag excess. However, the excess uptake does
not exceed about 30 \% of the symmetric \ensuremath{\sum}5 (210) GB's
excess. This does not necessarily imply a generally fixed saturation
level, as it depends on the local chemical potential, which is influenced
also by the annealing temperature and subsequent cooling/quenching
conditions. In light of an extrapolation of the facet data to FSR
= 0 in Figure 4b }\textit{\textcolor{black}{\emph{(white dashed line)}}}\textcolor{black}{,
the asymmetric GB should have a }\textit{\textcolor{black}{\emph{Ag}}}\textcolor{black}{{}
concentration of around 16 \% of a \ensuremath{\sum}5 (210) boundary.
As this is less than the actually observed 30 \%, the asymmetric GB
seems much more favorable to be preserved compared to inducing a GB
dissociation. This agrees well with our previous MD simulation results,
which indicate Ag solute excess being absorbed at the asymmetric boundary
already starting at a chemical potential around $\mu=0.35\,eV$, while
the symmetric \ensuremath{\sum}5 (210) boundary starts taking up Ag
only above chemical potentials of $\mu=0.5\,eV$ \cite{peter2018segregation}.
Thus, Ag uptake of the asymmetric boundary before the threshold concentration
seems to be a prerequisite to induce the nanofaceting. Finally, the
jump from about 30 \% relative Ag monolayer coverage (FSR = 0) to
about 50 \% when nanofaceting is induced, strongly indicates a first
order GB transition occurring. }

Our experimental findings enable us to propose a mechanistic picture
of the observed faceting phenomenon. We showed previously experimentally
and by simulations that heat treatment at 800°C (1073 K) does not
induce nanofaceting at the asymmetric GB studied here, while Ag segregation
does. Consequently, heating the boundary to 800ºC leaves the boundary
intact, possibly with increased disorder. Mishin and co-workers showed
in a series of studies on Cu-Ag GBs that increasing temperature, but
also increasing Ag solute excess can significantly influence how ordered
a GB is referred to the perfect bulk lattice \cite{williams2009thermodynamics,suzuki2005atomic,koju2020relationship,hickman2016disjoining}.
At such high temperatures, Ag atoms are going almost completely in
solid solution and diffuse from the surface through the bulk and along
the boundary creating a Ag concentration gradient. The solidus line
was likely not crossed, as no homogeneous nucleation of Ag precipitates
was observed inside the grains. Upon cooling, the disordered grain
boundary restores its low temperature structure, while still remaining
in the solid solubility region according to the equilibrium phase
diagram \cite{subramanian1993ag}. Judging from the determined bulk
Ag concentration of about 1.5 at. \%, the solidus line is crossed
at around 550°C (823 K). At this point, Ag starts to segregate towards
the GB upon further cooling to minimize the energy of the system.
The amount of Ag segregating to the boundary is depth dependent due
to the created concentration gradient. Eventually, a critical threshold
concentration is reached to initiate nucleation of the symmetric \textcolor{black}{\ensuremath{\sum}5
(210) facets, since the observed normal kite structure (Fig. 2, 2-4)
can absorb the highest amount of Ag excess and leads to the largest
reduction in energy compared to segregation to the initial asymmetric
GB \cite{peter2018segregation}. The nucleation of symmetric Ag-rich
facets is the initial step dissociating the asymmetric GB into a nanofaceted
boundary. The local amount of Ag solute excess at the boundary, which
is dictated by the concentration gradient in the bulk, determines
the density of nucleation sites and provides the driving force for
facet growth. Following the trend of having more Ag towards the surface,
symmetric facets grow longer towards the surface until a continuous
\ensuremath{\sum}5 (210) GB is formed close to the surface. Below
an estimated critical Ag threshold concentration of around 0.6 at.\%
for the bulk Ag content faceting will not occur considering a constant
segregation factor ($\beta=\frac{GBconcentration}{Bulkconcentration})$.
However, as the bulk diffusion gradient cannot be quantified here
accurately, the threshold concentration cannot be determined precisely
neither. The nucleation of the Ag-rich \ensuremath{\sum}5 (210) facet
segments at the initially asymmetric GB is accompanied by the nucleation
of the asymmetric facet segment and the associated GB disconnections
or facet junctions, which are also necessary to migrate the boundary
to compensate for the overall change in inclination towards the surface.
The defects are likely involved in the migration of the GB to compensate
for the inclination change induced by nucleation the symmetric facet
segments. For instance, the migration of a flat, symmetric GB was
shown theoretically for an Al \ensuremath{\sum}7 GB to function via
a nucleation process of islands on the grain boundary plane \cite{hadian2016atomistic}
(also regarded as disconnection nucleation for \ensuremath{\sum}13
and \ensuremath{\sum}17 Cu boundaries \cite{combe2016disconnections,rajabzadeh2013elementary})
and to be highly temperature dependent between 500 K to 700 K \cite{hadian2016atomistic},
which is close or even in the temperature range we expect our facet
nucleation to occur. The question remains, why a nanofaceted GB portion
is found instead of coarse facets or even just a curved GB. In general,
the energy of faceted GB following Hamilton et al. is composed of
two parts: (i) the contribution of the GB energies of the two facet
segments $\gamma_{1,2}$ along with the contribution of facet junctions
$\gamma_{j}$ as well as (ii) the interaction between facet junctions
$\gamma_{j-j}$. The overall GB energy $\gamma_{GB}$ can thus be
described as $\gamma_{GB}=\gamma_{1}+\gamma_{2}+\frac{2\gamma_{j}}{\varLambda}+\frac{\gamma_{j-j}}{\Lambda}$,
with $\varLambda$ being the facet period (length). For our purposes
we neglect the interaction of the junction term here leading to $\gamma_{GB}=\gamma_{1}+\gamma_{2}+\frac{2\gamma_{j}}{\varLambda}$.
Hamilton et al. demonstrated by simulation work that the GB inherently
wants to create equal sized, coarse facets to reduce the spacing of
the GB facet junctions and thus $\gamma_{GB}$ as much as possible
\cite{hamilton2003grain}. Although this theory was confirmed experimentally
by Medlin et al. \cite{medlin2003accommodation} and atomistic simulations
by Wu et al. \cite{wu2009grain}, this analysis does not consider
segregation and kinetic effects. Wu et al. found that migration kinetics
during facet coarsening can promote the formation of finite sized
facets. Instead of a continuous decrease in energy for increasing
facet periods, migration of facet segments can impose a multitude
of energy barriers depending on the structure of the facet segment
resulting in the inhibition of further migration and therefore stabilization
of finite length facets. Here, the energy change due to segregation
can be expressed generally through $\Delta\gamma_{GB}=\Delta\gamma_{ASL}+\Delta\gamma_{SSL}+\frac{2\Delta\gamma_{j}}{\varLambda}$.
However, considering the preferential segregation only to the symmetric
portion and no noticeable segregation to facet segment junctions,
the energy reduction is mainly attributed to the reduction in energy
due to the formation of symmetric \ensuremath{\sum}5 (210) segments
thus following $\Delta\gamma_{GB}=\Delta\gamma_{2}$. Nevertheless,
it would be interesting to study effects of segregation to junctions
and how they influence the energetics of the systems including the
interaction between junctions, which was neglected here. Segregation
to facet junctions was for example recently observed at faceted Si
\cite{liebscher2018strain} and Cu GBs \cite{meiners2020tantalum}. }

In conclusion, we present the first GB faceting diagram by studying
the influence of Ag solute excess concentration on facet formation
in an asymmetric tilt grain boundary via aberration corrected STEM,
STEM-EDS and APT. After having established a Ag concentration gradient
by a diffusion couple approach, we were able to distinguish different
structural arrangements (faceting states) along the boundary as a
function of Ag solute excess concentration. \textit{\textcolor{black}{\emph{The
total length of the \ensuremath{\sum}5 GB increases by more than 15\%
compared to its initial (Ag-free) position, revealing a lower energy
state after Ag segregation.}}} In the bulk at low Ag solute excess
relative to the surface, the initially asymmetric GB remains structurally
unaffected but takes up Ag. Exceeding a threshold amount of solute
excess, the boundary undergoes a first order transition and becomes
nanofaceted. One facet segment\textcolor{red}{\emph{ }}\textit{\textcolor{black}{\emph{changes
its grain boundary plane, but}}} remains asymmetric, Ag-lean and constant
in length, while the other segment was determined to be a symmetric
\textcolor{black}{\ensuremath{\sum}5 (210) type, with increasing length
towards the surface and being enriched in Ag. At the surface, highest
amounts of available Ag allow for a planar \ensuremath{\sum}5 (210)
GB containing close to a monolayer of Ag. }Similar GB phase diagrams
have been established for different transitions like GB premelting,
intergranular film formation or for multilayer segregation patterns
as a function of concentration \cite{cantwell2020grain}. However,
only few such GB phase diagrams could be established, yet. We present
here the first experimental GB nanofaceting diagram as a function
of solute excess concentration, similar to GB segregation diagrams
as proposed by Lejček and Hofman for iron GBs \cite{lejvcek1993grain}
or Watanabe et al.'s hardening/segregation GB diagrams in iron alloys
\cite{watanabe1980grain}, but focusing on the atomic grain boundary
structure evolution. Establishing such diagrams, analogous to bulk
phase diagrams, may offer new possibilities for microstructure engineering
in order to tune macroscopic materials properties. Although the preferential
segregation might appear as an exception from the Langmuir McLean
adsorption behavior in terms of homogeneous GB decoration, the individual
segments still follow Langmuir McLean. \textcolor{black}{Regarding
the asymmetric segregated and the symmetric \ensuremath{\sum}5 (210)
GB, we were able to use the Langmuir McLean relation to deduce the
ratio of their respective segregation energies and found that the
Gibbs segregation energy of the symmetric boundary is approximately
1.08 times the energy of the asymmetric GB.} Nevertheless, adjusted
segregation models accounting for such faceting transitions are becoming
necessary for these cases. \textcolor{black}{It has to be noted that
a continuously curved grain boundary, in contrast to a faceted boundary
may be connected differently to physical properties such as mobility
or plastic deformation. In fact, continuous curvatures at lower magnifications
might be related to nanofaceting transitions as observed in the present
study more often for doped material systems. Still, such phenomena
would only be resolved with adequate characterization techniques of
highest resolution. }

\section*{Methods}

\textbf{\textcolor{black}{Bicrystal growth and sample preparation.}}\textcolor{black}{{}
To achieve the structure-chemistry correlation for Ag segregation
induced nanofaceting, we grew a Cu bicrystal containing the asymmetric
grain boundary using the Bridgman technique. The {[}001{]} crystallographic
directions of the two seed crystals were aligned parallel with a misorientation
of 54\ensuremath{^\circ} around this common axis to ensure the formation
of a tilt GB with a common {[}001{]} zone axis orientation, important
for the STEM analysis. Additionally, an asymmetric inclination of
31\ensuremath{^\circ} was chosen for the as-grown reference boundary.
From the grown bicrystal, discs were cut perpendicular to the {[}001{]}
growth direction and metallographically prepared to obtain best surface
quality with a final electropolishing step. A reference lamella was
extracted from this disc for STEM investigations using focused ion
beam (FIB) preparation. Segregation was achieved by thermal annealing
at 800\ensuremath{^\circ}C under high vacuum conditions for 120 hours.
In our previous work we could exclude already that these conditions
influence the pure, Ag-free, grain boundary structure \cite{peter2018segregation}.
Ag was sputter deposited to a thickness of 500 nm onto the same disc
used as reference, to extract a Ag segregated GB as close to the reference
state as possible. In this way a diffusion couple arrangement was
created between Ag and the GB-containing Cu. No additional metallographic
preparation was performed to not alter the state of the GB artificially.
Thus, although a slight spatial misalignment, we believe to examine
the same GB before and after the diffusion process. }

\textbf{\textcolor{black}{Scanning transmission electron microscopy.}}\textcolor{black}{{}
The FIB prepared lamella was investigated after thermal treatment
using a probe-corrected STEM in a Thermo Fischer Scientific Titan
Themis 60-300 STEM machine. A schematic overview of the crystallography
and diffusion couple geometry is provided in Fig. 1(a). The shown
micrographs were acquired in high-angle annular dark field imaging
conditions to benefit from its Z-contrast conditions, i.e. the semi-collection
angle was set to 76 - 200 mrad. EDS maps were acquired at selected
locations to correlate nanofaceting with Ag solute excess concentration
in zone axis orientation immediately after image acquisition to avoid
tilting and thus loss of spatial correlation. A Super-X windowless
EDS detector was used at an acceleration voltage of 300 kV. }

\textbf{\textcolor{black}{Atom probe tomography.}}\textcolor{black}{{}
Quantification of Ag excess concentration at the GB with respect to
the matrix was obtained from atom probe tomography (APT) reconstructions
at specific regions of interest. APT tips were prepared by FIB (same
instrument used for TEM lamellae) \cite{thompson2007situ}, applying
a final cleaning at 5 kV and 41 pA, ensuring a Ga content <0.1 at.\%
in the analyzed regions. APT data was collected using a local electrode
atom probe, LEAP 5000XS by Cameca Instruments. The measurements were
performed at a temperature of 50 K, in laser mode with an energy of
60 pJ, using a pulse repetition rate of 250 kHz and a detection rate
of 0.007 atoms per pulse.}

\section*{Acknowledgments}

The authors cordially thank Tobias Oellers from the Ruhr University
Bochum for Ag deposition on our Cu bicrystals to create the investigated
diffusion couples. The funding and support by the European Research
Council (ERC) under the EU’s Horizon 2020 Research and Innovation
Programme is gratefully acknowledged by NJP, MJD and GD (\textcolor{black}{ERC
Advanced Grant, GB-Correlate,}\textit{\textcolor{red}{{} }}Grant No.
787446).

\section*{References}

\bibliographystyle{naturemag}

\begin{figure*}[!htbp]
\centering \includegraphics{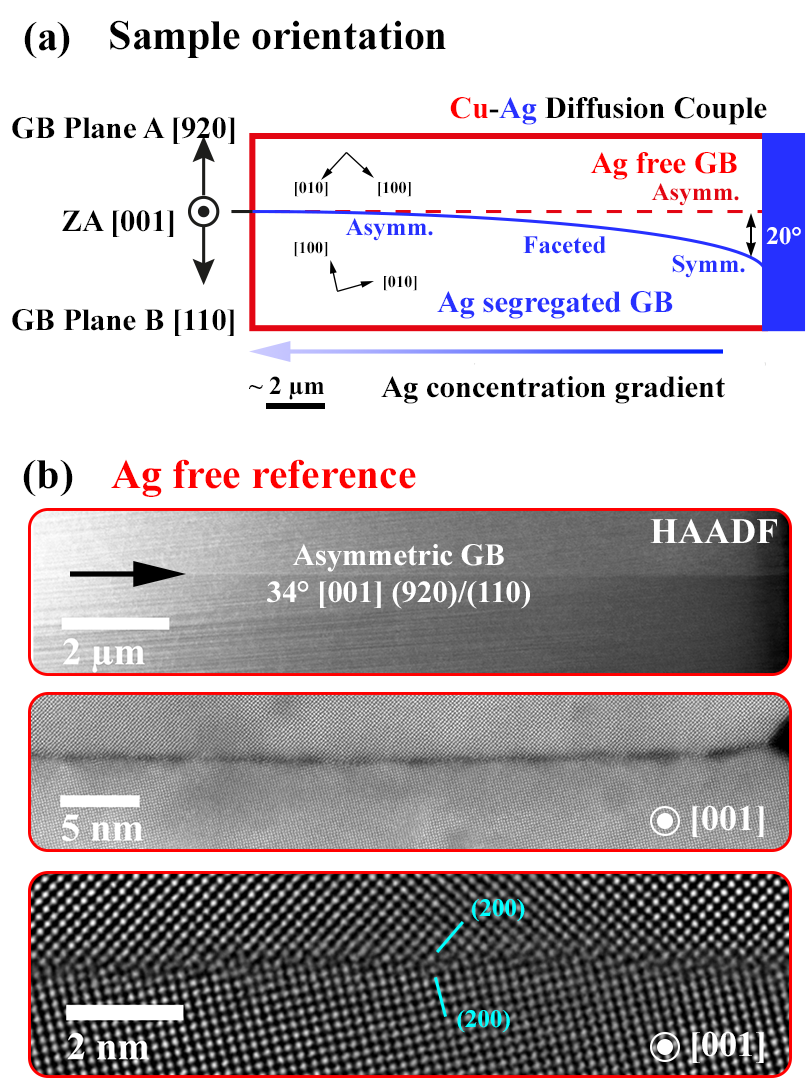}\caption{(a) Schematic representation of the diffusion couple's arrangement
and the crystallography of the grown bicrystal containing the asymmetric
tilt grain boundary and how the boundary behaves after segregation.
(b) HAADF-STEM micrographs of the asymmetric, Ag-free, reference boundary
at different magnifications. Micrographs are in {[}001{]} zone axis
orientation.}
\end{figure*}

\begin{figure*}[!htbp]
\centering 

\includegraphics{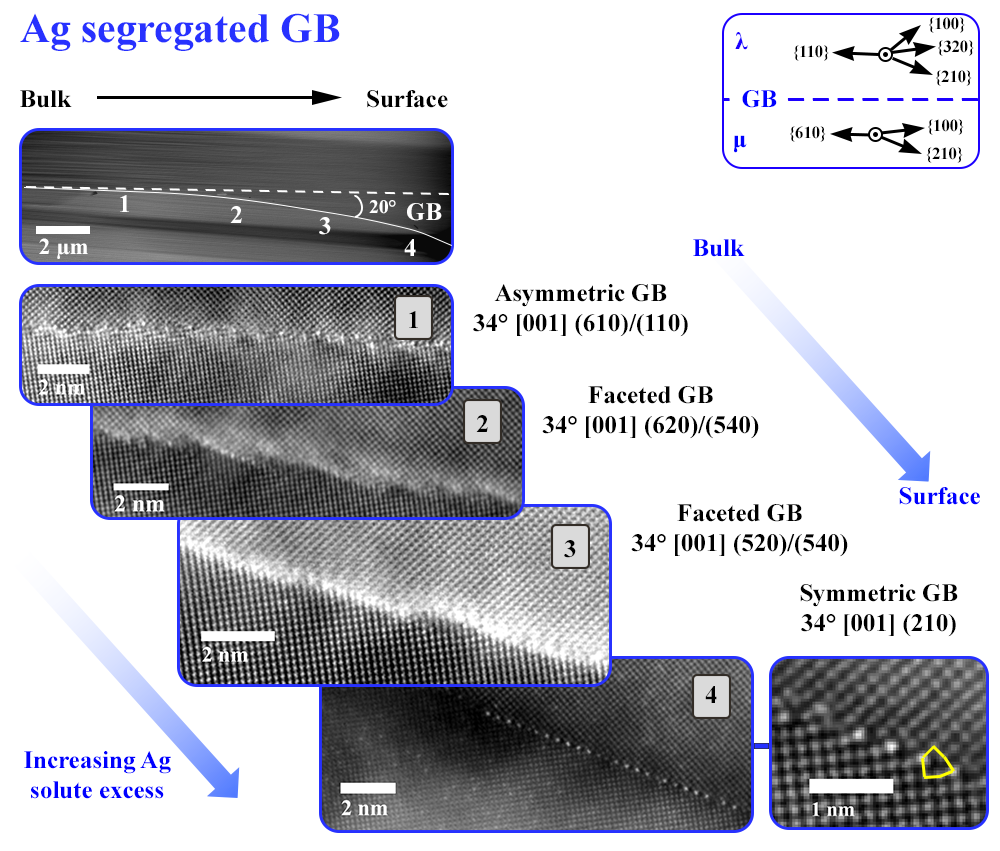}\caption{The curved Ag segregated GB as an overview image at lower magnification
along with micrographs along a boundary over a distance of about 15
\textmu m indicating the changes from an asymmetric GB (bulk) over
the nanofaceted state with increasing symmetric facet segments to
a solely symmetric GB (surface) and an atomic resolution micrograph
to capture the symmetric (210) GB's kite structure. All HAADF-STEM
micrographs were acquired in {[}001{]} zone axis orientation.}
\end{figure*}

\begin{figure*}[!htbp]
\centering 

\includegraphics{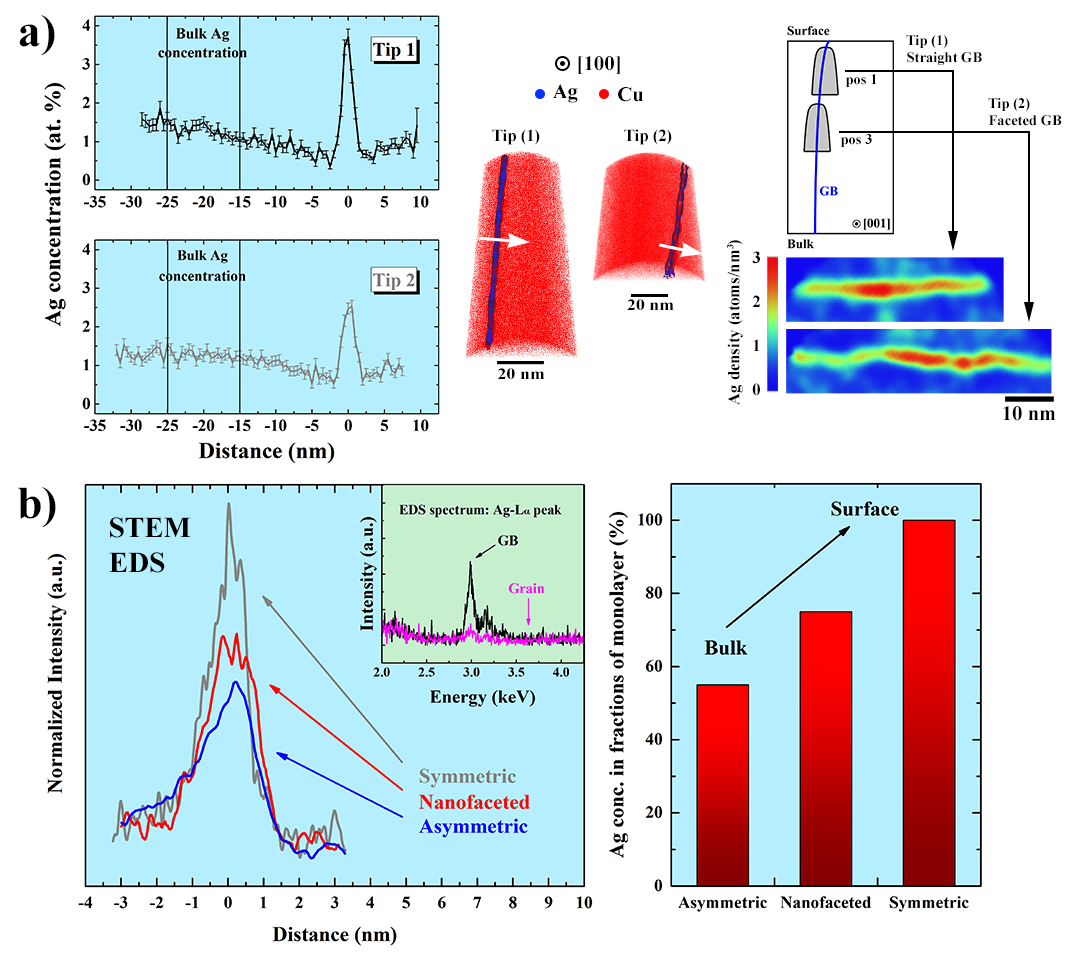}\caption{(a) APT results of the Ag segregated GB. The extracted concentration
profile across the boundary is given, while Ag density maps are shown
representative for two different samples - one at the surface, one
deeper into the bulk - highlighting the different GB states. (b) STEM-EDS
results across the GB at three representative locations. In the nanofaceted
case, EDS information is collected across the GB containing multiple
facets. EDS data was collected in zone axis orientation at the exact
location where micrographs were recorded. }
\end{figure*}

\begin{figure*}[!htbp]
\centering 

\includegraphics{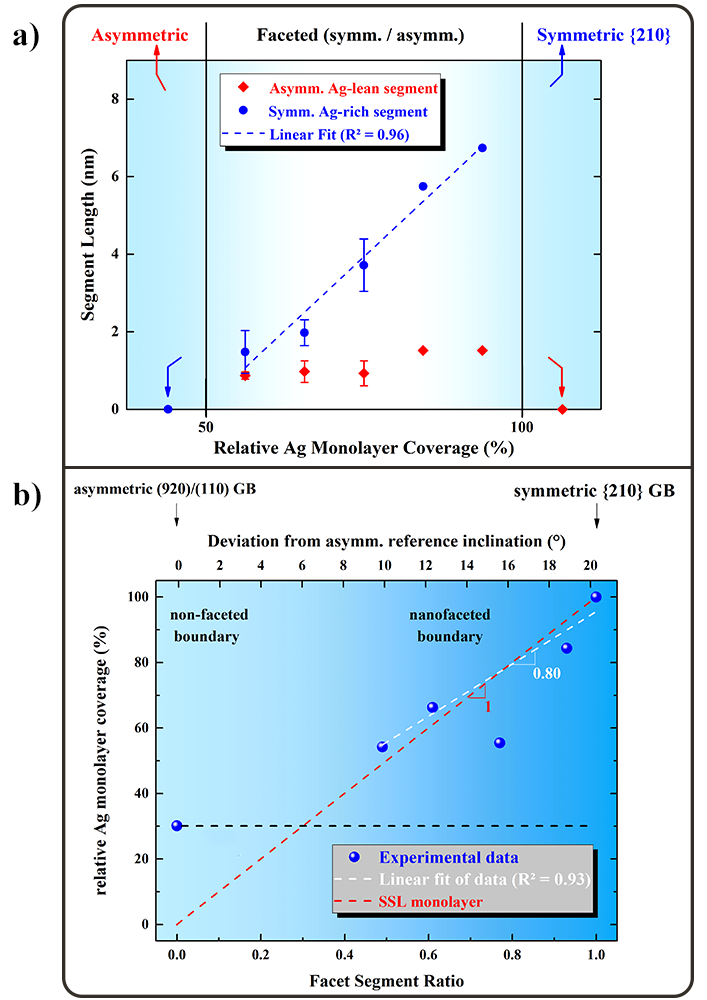}\caption{(a) GB faceting diagram presenting the facet segment length evolution
along the boundary as a function of Ag solute excess concentration.
Three distinct GB states appear: (i) asymmetric and not faceted for
low Ag excess, (ii) nanofaceted, preferentially segregated boundary
within a solute excess window and (iii) symmetric and not faceted
boundary for high Ag excess. (b) Correlation between faceting, inclination
and overall inclination of the GB including a linear fit through recorded
data points (white line), a straight line to highlight an expected
linear dependence of Ag content in facets (red) and the boundary specific
Ag excess at the asymmetric GB.}
\end{figure*}

\end{document}